\documentstyle[12pt]{article}
\input{epsf}

\newlength{\minitwocolumn}
\catcode`\@=11
\def\section{\@startsection {section}{1}{\z@}{-3.5ex plus -1ex minus 
    -.2ex}{2.3ex plus .2ex}{\bf }}

\newbox\tempboxa
\newdimen\captionboxsubcount 
\def\capsize#1{\captionboxsubcount=#1pt}
\newdimen\captionboxsub
\captionboxsub=\hsize \advance\captionboxsub by -\captionboxsubcount
\advance\captionboxsub by -\captionboxsubcount
\long\def\@makecaption#1#2{
 \setbox\@tempboxa\hbox{{\footnotesize\baselineskip=8pt #1: #2}}
 \ifdim \wd\@tempboxa >\captionboxsub 
\rightskip=\captionboxsubcount \leftskip=\captionboxsubcount 
{\footnotesize\baselineskip=8pt #1: #2}
\else \hbox to\hsize{\hfil\box\@tempboxa\hfil} 
 \fi}
\catcode`\@=12
\capsize{30}

\begin{document}
\begin{flushright}
\begin{minipage}{4cm}
\begin{flushleft}
\baselineskip=14pt
SU-4240-655\\
hep-ph/9702332\\
February, 1997
\end{flushleft}
\end{minipage}
\end{flushright}
\begin{center}
\large\bf
Existence of sigma meson in pi-pi scattering\footnote{
Talk given at 1996 International Workshop on ``Perspectives of Strong
Coupling Gauge Theories'' (Nov. 13--16, 1996, Nagoya).
This talk is based on the work done in collaboration with
Prof.~J.~Schechter and Dr.~F.~Sannino in 
Ref.~\ref{ref: Harada-Sannino-Schechter}.%
\nocite{Harada-Sannino-Schechter}
}
\end{center}
\begin{center}
{
Masayasu {\sc Harada}\footnote{
e-mail: {\tt mharada@npac.syr.edu}}}
\\
{\small\it
Department of Physics, Syracuse University\\
Syracuse, NY 13244-1130, USA}
\end{center}
\begin{abstract}
\baselineskip=13pt
In this talk I summarize a recently proposed mechanism to understand
$\pi\pi$ scattering to 1 GeV.
The model is motivated by the $1/N_{\rm C}$ expansion to QCD, and
includes a current algebra contact term and resonant pole exchanges.
Chiral symmetry plays an important role in restricting the form of the
interactions.
The existence of a broad low energy scalar ($\sigma$) is indicated.
\end{abstract}

\section{Introduction}

The $\pi\pi$ scattering has been studied as an important test 
of the strong interaction.
Now QCD is known to be the fundamental theory of the strong
interaction. 
However, it is very difficult to reproduce the experimental data
directly from QCD.
One clue is given by the structure of the chiral symmetry,
which approximately exists in the QCD Lagrangian and
is broken by the strong interaction of QCD.
Another clue is given by the $1/N_{\rm C}$ expansion to QCD.
In the large $N_{\rm C}$ limit, 
QCD becomes a theory of weakly interacting mesons,
and 
the $\pi\pi$ scattering is expressed as an infinite sum of tree
diagrams of mesons.\cite{1/Nc}

The experimental data in 
the low energy region near $\pi\pi$ threshold can be
reproduced by using the information from
chiral symmetry.
This situation is easily understood by using a chiral Lagrangian
which includes pions only.
In addition, by including the higher derivative terms together with
one-loop effects, the applicable energy region is enlarged.
This systematic low energy expansion is called the chiral perturbation
theory.\cite{ChPT}

In the higher energy region, however,
the one-loop amplitude of chiral perturbation theory
violates the unitarity bound around 
$400-500$\,MeV in the $I=0$ $S$-channel.\cite{Gasser-Meissner}
For the $P$-wave amplitude,
we have the $\rho$ meson, and chiral perturbation theory
may break down at the resonance position.
The explicit inclusion of resonances in the high energy region
easily reproduces the amplitude, of course.

When we apply the large $N_{\rm C}$ argument to the practical $\pi\pi$
scattering, 
we cannot actually include an infinite number of resonances.
Moreover, the forms of interactions are not fully determined in
the large $N_{\rm C}$ limit.
Nevertheless, some encouraging features were previously found in an
approach which truncated the particles appearing in the effective
Lagrangian  to those with masses up to an energy slightly greater than
the range of interest.\cite{Sannino-Schechter}
Moreover, the chiral symmetry played an important role to restrict the
form of interaction, i.e., the effective Lagrangian was constructed by
using the information of chiral symmetry.
This seems reasonable phenomenologically and is what one
usually 
does in setting up an effective Lagrangian. 

In this talk I concentrate on the energy region below 1\,GeV.
For the established resonances lighter than 1\,GeV,
$\rho$ and $f_0(980)$ are contained in the particle data group (PDG)
list\cite{ParticleDataGroup:94} (see Table~\ref{tab: particles}).
\begin{table}
\begin{center}
\begin{tabular}{|c||c||c|c|} \hline
 &$I^G(J^{PC})$ & $M$(MeV) & $\Gamma$(MeV) \\ 
\hline \hline
$\sigma(400-1200)$ & $0^+(0^{++})$  & 400$-$1200 & 600$-$1000 \\
$\rho(770)$   & $1^+(1^{--})$  & 769.9   & 151.2  \\
$f_0(980)$   & $0^+(0^{++})$  & 980     & 40$-$100 \\
\hline
\end{tabular}
\end{center}
\caption[]{
Resonances included in the $\pi\pi\rightarrow\pi\pi$ channel as
listed in the PDG. 
}
\label{tab: particles}
\end{table}
However, the width of $f_0(980)$ is not well determined.
Moreover, the existence of a light scalar $\sigma$ is suggested by
several authors.\cite{sigma}
Here I will determine these resonance parameters by fitting to the
$I=0$ $S$-wave $\pi\pi$ scattering amplitude.

This paper is organized as follows.
In section~\ref{sec: model} I will briefly show the interaction terms.
Section~\ref{sec: fit} is the main part of this talk, where I will
show how to regularize the amplitude, and fit the resonance
parameters to the experimental data of the $I=0$, $J=0$ partial wave
amplitude.
Finally, a summary is given in section~\ref{sec: summary}.

\section{Interaction Terms}
\label{sec: model}

In this section
I will briefly show the interactions between resonances and pions.
First I include the vector meson as a gauge field of chiral
symmetry\cite{Kaymakcalan-Schechter},
which is equivalent to the hidden local gauge 
method
(See, for a review, Ref.~\ref{ref: Bando-Kugo-Yamawaki:PRep}.)
\nocite{Bando-Kugo-Yamawaki:PRep}
at tree level.
The $\rho\pi\pi$ interaction is given by
\begin{equation}
{\cal L}_\rho = g_{\rho\pi\pi} \vec{\rho}_\mu \cdot
\left( \partial^\mu \vec{\pi} \times \vec{\pi} \right)
\ ,
\end{equation}
where $g_{\rho\pi\pi}$ is the $\rho\pi\pi$ coupling constant.
Next, I include scalar resonances, $\sigma$ and $f_0(980)$.
These are iso-singlet fields; the interaction with two pions is
given by
\begin{equation}
{\cal L}_{f}=-\frac{\gamma_f}{\sqrt{2}}\; f\;
\partial_{\mu}\vec{\pi}\cdot\partial_{\mu}\vec{\pi}
\qquad \left( f=\sigma\,,f_0(980) \right)
\ .
\label{la:sigma}
\end{equation}
Here I should note that the
chiral symmetry requires derivative-type interactions
between the scalar field and pseudoscalar mesons.

\section{Fit to $\pi\pi$ scattering to 1 GeV}
\label{sec: fit}

In this section, I will calculate the $S$-wave $\pi\pi$ scattering
amplitude by including resonances as explained in the previous
section.

The most problematic feature involved in comparing the leading
$1/N_{\rm C}$ amplitude
with experiment is that it does not satisfy unitarity.
Since the mesons have zero width in the large $N_{\rm C}$ limit,
the amplitude diverges at the resonance position.
Thus in order to compare the $1/N_{\rm C}$ amplitude with experiment
we need to regularize the resonance contribution.
The ordinary narrow resonances such as $\rho$ meson are regularized by
including the width in the denominator of the propagator
(the Breit-Wigner form):
\begin{equation}
\frac{M\Gamma}{M^2-s-iM\Gamma}\ .
\label{Breit-Wigner}
\end{equation}
This is only valid for a narrow resonance in a region where the
{\it background} is negligible.
Note that the width in the denominator is related to the coupling
constant.

For a very broad resonance
there is no guarantee that such a form is correct. 
A suitable form turned out to be of the type
\begin{equation}
\frac{M G}{M^2-s-iMG^\prime}\ ,
\label{sigma-propagator}
\end{equation}
where the parameter $G^\prime$ is a free
parameter which is not related to the coupling constant.

Even if the resonance is narrow, the effect
of the background may be rather important.
This seems to be true for the case of the $f_0(980)$.
Demanding local unitarity in this case yields a partial
wave amplitude of the well known form:\cite{Taylor}
\begin{equation}
\frac{e^{2i\delta}M\Gamma}{M^2-s-iM\Gamma}+e^{i\delta}\sin \delta\ ,
\label{rescattering}
\end{equation}
where $\delta$ is a background phase (assumed to be slowly varying).
I will adopt a point of view in which this form is regarded as a kind
of regularization of the model. Of course, non zero $\delta$
represents a rescattering effect which is of higher order in
$1/N_{\rm C}$.
The quantity $\displaystyle{e^{2i\delta}}$, taking $\delta=constant$,
can be incorporated into the squared coupling constant connecting the
resonance to two pions.
In this way, crossing symmetry can be preserved.
The non-pole background term in Eq.~(\ref{rescattering}) and hence
$\delta$ is to be predicted by the other pieces in the effective
Lagrangian. 

Another point which must be addressed in comparing the leading
$1/N_{\rm C}$ amplitude
with experiment is that it is purely real away from the singularities.
The regularizations mentioned above do introduce some imaginary pieces
but these are clearly more model dependent.
Thus it seems reasonable to compare the real part of the predicted
amplitude with the real part of the experimental amplitude.

Let me start from the {\it current algebra} + $\rho$ contribution.
The predicted curve is shown in Fig.~1 of 
Ref.~\ref{ref: Harada-Sannino-Schechter}.
Although the introduction of $\rho$ dramatically
improves unitarity up to about $2$\,GeV, $R^0_0$ violates unitarity to
a lesser extent starting around $500$\,MeV.
To recover unitarity, we need a negative contribution to the real part
above this point, while below this point a positive contribution is
preferred by experiment.
Such behavior matches with the real part of a typical
resonance contribution.
The resonance contribution is positive in the energy region below its
mass, while it is negative in the energy region above its mass.
Then I include a low mass broad scalar resonance, $\sigma$.
The $\sigma$ contribution to the real part of the amplitude component
$A(s,t,u)$ is given by
\begin{equation}
A_{\sigma}(s,t,u)=
\frac{\gamma_\sigma^2}{2}
\frac{(s-2m_\pi^2)^2}{M_\sigma^2-s - i M_\sigma{G^\prime}}\ ,
\label{eq:sigma}
\end{equation}
where the factor $(s-2m_{\pi}^2)^2$ is due to the derivative-type
coupling required for chiral symmetry in Eq.~(\ref{la:sigma}).
$G^{\prime}$ is a parameter which we introduce to regularize the
propagator.
It can be called a width, but it turns out to be rather large so that,
after the $\rho$ and $\pi$ contributions are taken into account, the
partial wave amplitude $R^0_0$ does not clearly display the
characteristic resonant behavior.

A best overall fit is obtained with the parameter
choices; $M_{\sigma}=559$\,MeV, 
$\gamma_{\sigma}=7.8$\,GeV$^{-1}$ and $G^{\prime}=370$\,MeV.
The result for the
real part $R^0_0$ due to the inclusion of the 
$\sigma$ contribution along
with the $\pi$ and $\rho$ contributions is shown in Fig.~1.
It is seen that the unitarity bound is satisfied and there is a
reasonable agreement with the experimental 
points\cite{Alekseeva,Grayer} up to about $800$\,MeV.

\begin{figure}[htbp]
\noindent
\begin{minipage}[t]{\minitwocolumn}
\begin{center}
\epsfxsize=0.92\minitwocolumn
\ \epsfbox{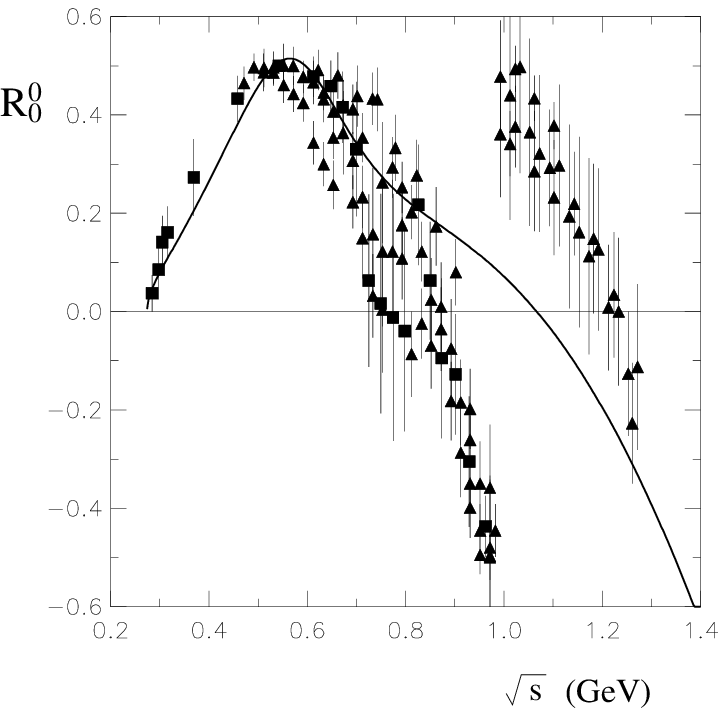}
\end{center}
\addtocounter{figure}{1}
\noindent
\footnotesize
Figure~\thefigure:
\label{figura2}
The solid line is the {\it current algebra}
$+~\rho+\sigma$ result for $R^0_0$.
The experimental points, in this
and succeeding figures, are extracted from the phase shifts
($\Box$: Ref.~\ref{ref: Alekseeva}\nocite{Alekseeva},
$\triangle$: Ref.~\ref{ref: Grayer}.\nocite{Grayer}).
\end{minipage}
\hspace{\columnsep}
\begin{minipage}[t]{\minitwocolumn}
\begin{center}
\epsfxsize=0.92\minitwocolumn
\ \epsfbox{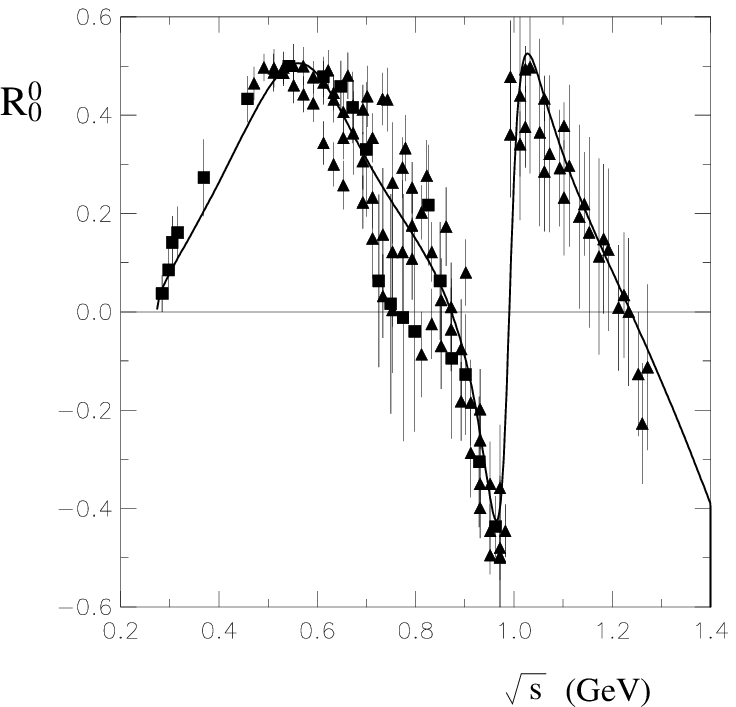}
\end{center}
\addtocounter{figure}{1}
\noindent
\footnotesize
Figure~\thefigure:
The solid line is the {\it current algebra}
$+~\rho~+~\sigma~+~f_0(980)$ result for $R^0_0$ obtained by assuming
the values in Table~\ref{tab: result}
for the $\sigma$ and $f_0(980)$ parameters.
\label{figura4}
\end{minipage}
\end{figure}

Next, let me consider the 1 GeV region.
Reference to Fig.~1 shows that the experimental data for
$R^0_0$ lie considerably lower than the $\pi+\rho+\sigma$ contribution
between $0.9$ and $1.0$\,GeV and then quickly reverse sign above this
point.
This is caused by the existence of $f_0(980)$.
As we can see easily, a naive inclusion of $f_0(980)$ does not
reproduce the experimental data,
since the real part of the typical resonance form gives a positive
contribution in the energy region below its mass, while it gives a
negative contribution in the energy region above its mass.
However, we need a negative contribution below 1\,GeV
and a positive contribution above 1\,GeV.

As I discussed around Eq.~(\ref{rescattering}), the effect of the
background is important in this $f_0(980)$ region.
In this case the background is 
given by the $\pi+\rho+\sigma$ contribution.
Figure~1 shows that the real part of the background is
very small so that the background phase $\delta$ 
in Eq.~(\ref{rescattering}) is expected to be roughly 90$^\circ$.
This background effect generates the extra minus sign in front of the
$f_0(980)$ contribution, as we can see from Eq.~(\ref{rescattering}).
Thus the $f_0(980)$ gives a negative contribution below the resonance
position and gives a positive contribution above it.
This is clearly exactly what is needed to bring experiment
and theory into agreement up till about $1.2$\,GeV.

The actual amplitude used for the calculation properly
contains the effects of the pions' derivative coupling to the
$f_0(980)$:
\begin{equation}
A_{f_0(980)}(s,t,u) = e^{2i\delta} \frac{\gamma_{f_0\pi\pi}^2}{2}
\frac{(s-2m_\pi^2)^2}{M_{f_0}^2 - s - i M_{f_0} \Gamma_{f_0}}
\ ,
\end{equation}
where $\delta$ is a background phase parameter and the real coupling
constant $\gamma_{f_0\pi\pi}$ is related to the 
$f_0(980) \rightarrow \pi\pi$ width.
The background phase parameter $\delta$ is predicted by
\begin{equation}
\frac{1}{2} \sin(2\delta) \equiv \widetilde{R}_0^0(s=M_{f_0}^2)
\ ,
\end{equation}
where $\widetilde{R}_0^0$ is computed as the sum of the current
algebra, $\rho$, and sigma pieces.

A best fit of our parameters to the experimental data results in the
curve shown in Fig.~2.
Only the three parameters $\gamma_{\sigma}$, $G'$ 
and $M_\sigma$ are essentially free.
The others are restricted by experiment.
Since the total width of $f_0(980)$ has a large uncertainty 
(40 -- 100\,MeV in PDG list),
we also fit this.
In addition we have considered the precise value of $M_{f_0}$ to be a
parameter for fitting purpose.
The best fitted values are shown in Table~\ref{tab: result} together
with the predicted background phase $\delta$ and the $\chi^2$ value.
The predicted background phase is seen to be close to 90$^\circ$,
and the low lying sigma has a mass of around 560\,MeV and a width of
about 370\,MeV.
\begin{table}[hbtp]
\begin{center}
\begin{tabular}{|c|c|c|c|c|c|c|}
\hline
$M_{f_0(980)}$ & $\Gamma_{f_0(980)}$ & $M_\sigma$ & $G'$ 
 & $\gamma_{\sigma}$ & $\delta$ (deg.) & $\chi^2$ \\
\hline
987 & 64.6 & 559 & 370 & 7.8 & 85.2 & 2.0 \\
\hline
\end{tabular}
\end{center}
\caption[]{The best fitted values of the parameter together with the
predicted background phase $\delta$ and the $\chi^2$ value.
The units of $M_{f_0(980)}$, $\Gamma_{f_0(980)}$, $M_\sigma$ and $G'$
are MeV and that of $\gamma_{\sigma}$ is GeV$^{-1}$.
}
\label{tab: result}
\end{table}

\section{Summary}\label{sec: summary}

In this talk I showed the main mechanism of the analysis done in
Ref.~\ref{ref: Harada-Sannino-Schechter}:
(1) motivated by the large $N_{\rm C}$ approximation to QCD, we
include the resonances with masses up to an energy slightly greater
than the range of interest, and use the chiral symmetry to restrict
the forms of the interactions;
(2) the {\it current algebra} + $\rho$ contribution 
violates the unitarity bound
around 560\,MeV region but it is recovered by including the low mass
broad resonance sigma\cite{Sannino-Schechter};
(3) the $\pi$ + $\rho$ + $\sigma$ contribution gives an important
background effect to the $f_0(980)$ contribution, i.e., the sign in
front of the $f_0(980)$ contribution is reversed by the background
effect.
The third mechanism, which leads to a sharp dip in the $I=J=0$ partial
wave contribution to the $\pi\pi$-scattering cross section, can be
identified with the very old {\it Ramsauer-Townsend} effect 
\cite{Schiff} which concerned the scattering of $0.7~eV$ electrons on
rare gas atoms.
The dip occurs because the background phase of $\pi/2$ causes the
phase shift to go through $\pi$ (rather than $\pi/2$) at 
the resonance position.
(Of course, the cross section is proportional to 
$\sum_{I,J}^{~}(2J+1) \sin^2(\delta^J_I)$.)
This simple mechanism seems
to be all that is required to understand the main feature of $\pi\pi$
scattering in the $1~GeV$ region.

The detailed analysis, which includes the effects of the inelasticity
($\pi\pi\rightarrow K\overline{K}$ channel opens at 990\,MeV.) and the
next group of resonances, is done in
Ref.~\ref{ref: Harada-Sannino-Schechter}.
The results show that those effects only fine-tune the best fitted
values shown in Table~\ref{tab: result}.
Finally I should stress that the consistent neglect of the
crossed-channel $\rho$ exchange does not destroy the existence of
$\sigma$ meson.\cite{Harada-Sannino-Schechter:2}
The best fitted values are $M_\sigma=378$\,MeV and 
$G'=836$\,MeV.

\vspace{2pt}
\begin{flushleft}
\bf Acknowledgement
\end{flushleft}
\vspace{-5pt}

I would like to thank Prof.~Joe~Schechter and Dr.~Francesco~Sannino
for careful reading of this
manuscript, and the organizing committee of the conference for the
opportunity to present this talk.

\end{document}